\documentclass[11pt]{article}
\usepackage{graphicx}
\usepackage{cs12,html}
\begin{document}
 
\title{Modeling the corona of AB Doradus}
 
\author{G.A.J. Hussain\altaffilmark{1}, 
A.A. van Ballegooijen\altaffilmark{1}, 
M. Jardine\altaffilmark{2},
A. Collier Cameron\altaffilmark{2}}
\altaffiltext{1}{Harvard Smithsonian CfA}
\altaffiltext{2}{University of St Andrews}

\index{*AB Dor}
\index{3-D Simulations}

\begin{abstract}
We present a model for the coronal topology of the active, rapidly
rotating K0 dwarf, AB Doradus.  
Surface magnetic field maps obtained using a technique based on Zeeman
Doppler imaging indicate the presence of a strong non-potential component
near the pole of the star. The coronal topology is obtained by
extrapolating these surface maps. 
The temperature and density
in the corona are evaluated using an energy balance model.
Emission measure distributions computed using our models compare
favorably with observations. 
However, the density observed by EUVE, 
$n_e \approx 10^{13}$~cm$^{-3}$, 
at the emission measure peak temperature of 
$8 \times 10^6$~K remains difficult to explain satisfactorily.
\end{abstract}

\section{Introduction}

AB Doradus is a well studied example of the class of very active, rapidly
rotating cool stars that are just evolving onto the main sequence. Its
youth is indicated by strong lithium absorption and it rotates at 50 times
the solar rate (P$_{rot}$=0.51d). AB Dor's rapid rotation makes it an
ideal candidate for high resolution spectroscopic techniques such as
Doppler imaging.  Surface maps obtained using this technique since 1992
typically show the presence of a dark polar spot region which extends down
to about 70$^{\circ}$ latitude, co-existing with lower latitude spots near
the equator. Spots are recovered to within the resolution capability of
the techniques so it is likely that smaller spot features also exist at
the surface that cannot be resolved. Photometry of the star from 1978
indicates that AB Dor may follow a solar-type activity cycle spanning
22-23 years (Amado et al. 2001).

The locations of the lower-latitude spots change from year-to-year and are
believed to have lifetimes of about a month. The polar spot is much more
long-lived, as it has been recovered in every spot map obtained of AB Dor
except for the first, where there there is no evidence of a polar spot
({K\"urster} et al. 1997).  This early image was derived using data taken
in 1988 and its epoch coincides with a maximum in spot activity according
to the star's photometric light level. Hence the polar spot may be tied in
with the stellar activity cycle. The presence of the polar cap does
indicate that a considerable quantity of flux is emerging at the pole of
the star. This polar spot ``phenomenon'' is common to spot maps recovered
on other rapidly rotating systems ranging from RSCVn binaries to young G
dwarfs (Vogt \& Penrod 1983, Strassmeier 1990, Barnes et al. 1998).

\begin{figure}[h]
\hspace{2.5cm}
\includegraphics*[width=7cm]{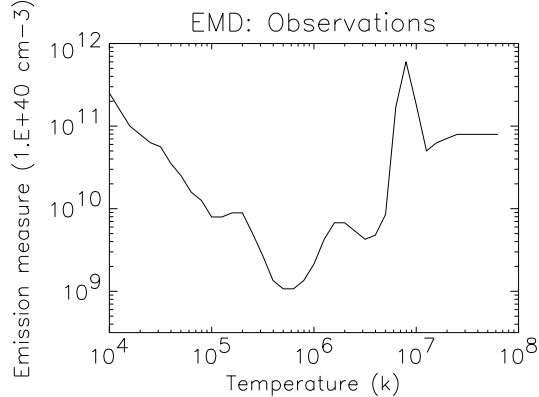}
\caption{Emission measure distribution from 1993-1994 for AB Dor 
computed from IUE and EUVE data (Sanz-Forcada et al. this volume). }
\label{fig:emdobs}
\end{figure}

\begin{figure}[h]
\hspace{2.5cm}
\includegraphics*[width=7.5cm]{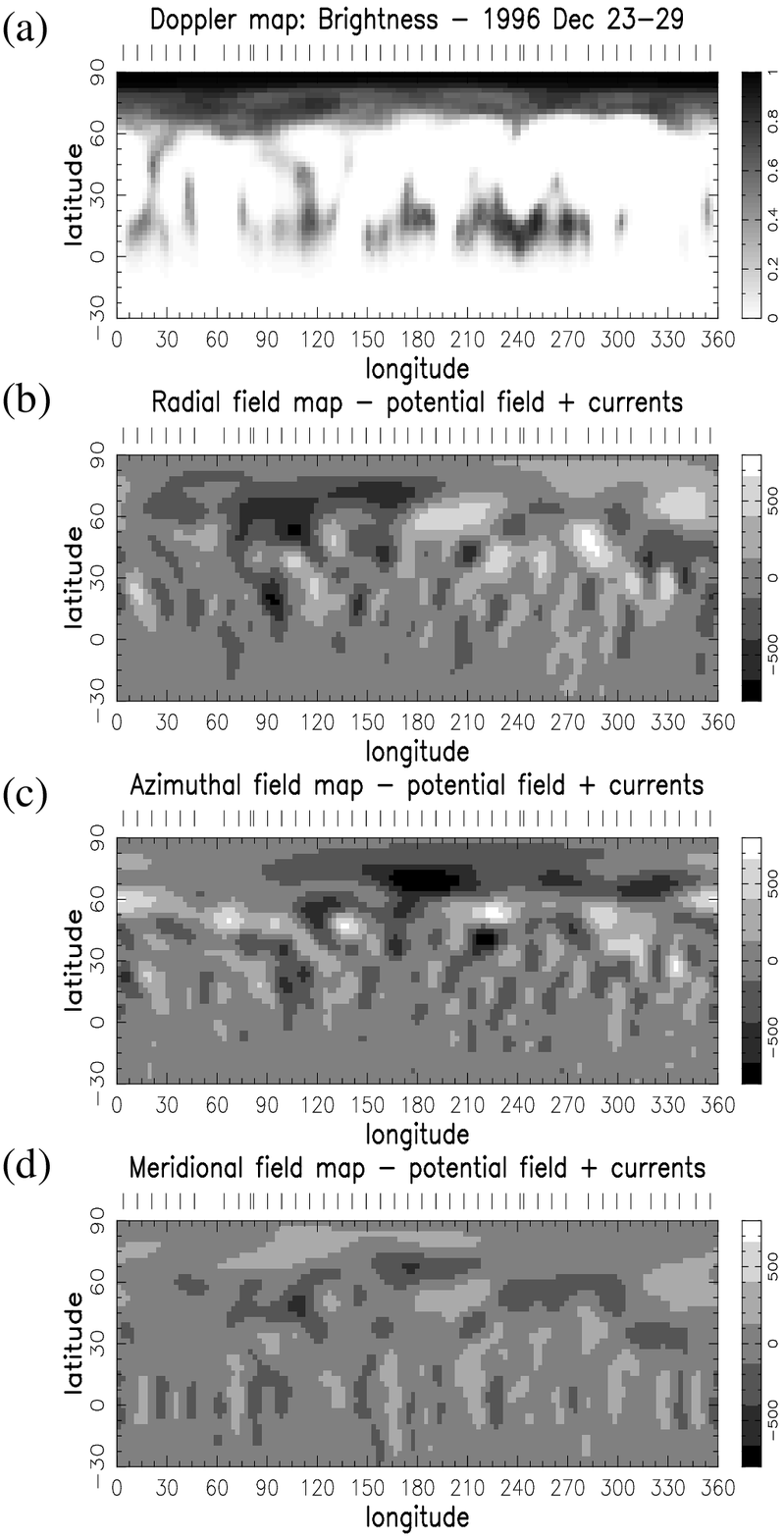}

\caption{Surface brightness map and magnetic field maps for AB Dor in 1996
Dec 23-29. The greyscale for the brightness map represents spot occupancy
for the star (1=complete spot coverage);  white/black represent $\pm$~800G
respectively in the magnetic field maps.  The magnetic field maps were
produced using the assumption that the electric current density, {\bf j},
in the corona is given by a potential field ($\mbox{\bf j}=-\nabla\Psi$).}
\label{fig:maps} 
\end{figure}

Coronal activity indicators in the X-ray, radio and optical wavelengths
suggest strong variability in the stellar corona. However, while it is
clear that AB Dor emits strongly in X-rays, 
L$_x$/L$_{bol} \approx 10^{-3}$ (Vilhu \& Linsky 1987) 
the characteristics of the emitting structures have yet to be ascertained.
 
Observations of the stellar corona obtained from a variety of sources
indicate the presence of both extended and compact structures. The
evidence for extended coronal features, with sizes of over a stellar
radius can be inferred from several sources.  Large prominence type
complexes are detected through fast-moving absorption transients moving
through the hydrogen Balmer lines.  These slingshot prominences are
supported at distances of between 2-5{\,\mbox{$\mbox{R}_*$}}\ and hence
well above the Keplerian corotation radius of the star 
($R_k \approx$ 2.6{\,\mbox{$\mbox{R}_*$}}) (Donati et al. 1999). 
The prominence lifetimes are not known but typically have upper limits of
about 2 rotation periods (i.e. 1 day). There is further evidence for
extended structures from FUSE and HST datasets. These reveal transition
region lines (C~{\sc iv}~1548\AA, Si~{\sc iv}~1393\AA, O~{\sc vi}~1032\AA)
with broadened wings in emission extending to velocities of 270km/s. These
wings are caused by optically thin plasma ($T \approx 10^5$~K) at heights
out to the co-rotation radius, 2.6{\,\mbox{$\mbox{R}_*$}}\ (Brandt et al.
2001, Young 2001).

Evidence for compact coronal features comes from flare decay analysis on
lightcurves obtained using BEPPOSAX (Maggio et al. 2000).  They find that
the flaring region is small ($H<0.3${\,\mbox{$\mbox{R}_*$}}) and that it
is uneclipsed over a full rotation cycle, indicating that it is located
near the pole of the star. A statistical analysis of ROSAT data spanning a
period of 5 years suggests rotational modulation at 5-13\% ({K\"urster} et
al. 1997). Brandt et al. (2001) find that between 60\% to 80\% of the
emission should originate near the stellar photosphere.

Measurements of pressure in the corona are extremely high, over two orders
of magnitude larger than those found on the quiet Sun. Using HST and
Chandra data, electron densities are found to be high, 
$n_e \approx 2-3 \times 10^{12}$~cm$^{-3}$, at transition region
temperatures (about $3\times 10^4$~K)  (Brandt et al. 2001, Linsky 2001).
At higher temperatures, around $3 \times 10^6$K, recent XMM results
indicate electron densities of $n_e \approx 3 \times 10^{10}$~cm$^{-3}$
({G\"udel} et al. 2001).  Sanz-Forcada et al. (this volume) compute an
emission measure distribution (EMD) for AB Dor by combining IUE and EUVE
observations taken in 1993 and 1994. The resulting EM is shown in
Fig.~\ref{fig:emdobs}. C~{\sc ii, iii, iv, v} lines from the IUE dataset
have been used to constrain emission measures under $\approx 10^5$K while
Fe~{\sc xix, xx, xxi, xxii} lines from the EUVE dataset have been used to
evaluate the emission measure at higher temperatures. Density sensitive
lines observed by EUVE indicate extremely high densities
($10^{12}<n_e<10^{13}$~cm$^{-3}$) at temperatures between $6-8\times
10^6$~K. Sanz-Forcada et al. (this volume) speculate that this EM peak is
caused by compact structures with strong magnetic fields that are capable
of supporting this kind of high-pressure gas. Results from similar rapidly
rotating systems indicate that this peak is very stable.  Brickhouse \&
Dupree (1998) suggest that an analogous EM peak observed in the EUVE EMD
of the contact binary, 44i~Boo, can be explained in terms of loops with
heights of about 0.004{\,\mbox{$\mbox{R}_*$}}. We set out to explain the
observations outlined in this section using simple models based on
extrapolations of AB Dor's surface magnetic field. We use magnetic field
maps obtained using the high resolution spectroscopic technique of Zeeman
Doppler imaging.

\section{Advanced ZDI}

Zeeman Doppler imaging (ZDI) is essentially Doppler imaging applied to
circularly polarized spectra (Semel 1989).  As with Doppler imaging, ZDI
relies on rotational broadening to separate the signatures from different
magnetic field regions on the surface in velocity-space.  Circularly
polarized spectra are sensitive to the {\em line-of-sight} component of
the magnetic field.  Within the weak field regime (under 1kG) the size of
the signature scales linearly with the size of the field.  By tracking the
velocity excursions and intensity variations of the circularly signatures
over the course of a rotation cycle we can determine the {\em orientation}
as well as the {\em size} of the surface field (Donati \& Brown 1997).  
Magnetic field maps of AB Dor obtained using this technique typically show
strong azimuthal field at the pole of the star. This very non-solar
pattern is puzzling and doubts have been cast over its authenticity,
partly due to uncertainties introduced in the conventional ZDI method.
Some uncertainties are due to the assumption that all field orientations
(radial, azimuthal and meridional fields) are assumed to be independent of
each other. This means that it is possible to reconstruct monopoles at the
stellar surface and equally physically unrealistic fields.

Hussain et al. (2001) describe an advanced version of ZDI that introduces
a relationship between different field orientations by assuming a
potential field. As well as assuming a relationship between radial,
azimuthal and meridional field vectors these maps can also be used to
extrapolate out to the stellar corona. The technique we present here is
more advanced as it incorporates departures from potential field
distributions. While this solution is not force-free it allows us to
pinpoint the areas where the surface field distribution becomes
non-potential while still allowing for the extrapolation of the field out
to the corona.

\subsection{Doppler maps}
The dataset used to obtain the maps shown in Fig.~\ref{fig:maps} was
obtained using the 3.9m Anglo-Australian telescope, UCL Echelle
spectrograph and the Semel polarimeter in 1996 December 23-29. Data were
reduced using the dedicated echelle reduction software package, {\sc
esprit} (Donati et al. 1997). Once reduced, the signal from over 1500
photospheric lines were ``summed up'' using a technique called {\em least
squares deconvolution}. This technique allows us to increase
signal-to-noise levels by a factor of 30. Full details on the instrument
setup, data reduction and deconvolution can be found in previous papers
(Donati et al. 1999, Hussain et al. 2000).

Data were combined from all four nights (1996 Dec. 23, 25, 27 and 29) in
order to obtain as complete phase coverage as possible. The surface
differential rotation rate as measured by Donati \& Collier Cameron (1997)
has been incorporated into the code to account for the shear of surface
features over the course of a week.  The brightness map was produced using
the Doppler imaging code, {\sc dots} (Collier Cameron 1997), and the
magnetic field maps shown here are produced using the code described
above. We tried initially to fit the observed circularly polarized spectra
using just a potential field distribution. However, it was not possible to
find a solution within a reduced $\chi^2 \approx 1.5$ for this dataset.  
We then tried a model in which the curl of the magnetic field
$\mbox{ \bf j} = \nabla \times \mbox{\bf B}$ is given by a potential field
($\mbox{\bf j}=-\nabla \Psi$, with $\nabla^{2} \Psi = 0$). The magnetic
field is determined by two free functions, $B_r(\theta,\phi)$ and
$j_r(\theta,\phi)$, at the stellar surface. This provides the additional
freedom to obtain a good fit to the polarization profiles and also allows
the magnetic field to be extrapolated into the corona in a realistic way.
The brightness and magnetic field maps are produced independently of each
other and are shown in Fig.~\ref{fig:maps}.

We find that the non-potential part of the magnetic field is found mostly
in the strong azimuthal field band above about 70$^{\circ}$ latitude. This
non-potential region is located around the boundary of the dark polar spot
that is recovered using Doppler imaging (c.f. Figs.~\ref{fig:maps}a~\&~c).  
The magnetic field distribution recovered using this method agrees with
the maps obtained previously using conventional ZDI (Donati et al. 1999).
The main differences between the ZD maps and these images are in the
azimuthal maps; specifically in the unbroken band of uni-directional flux.
While there is still a strong clockwise azimuthal component near the pole
in this model, it is weaker than that found in the Zeeman Doppler map and
is even broken by a region of opposite polarity. The strong azimuthal
field pattern is clearly non-solar.  On the Sun, horizontal field is
largely confined to the penumbrae of starspots and would cover a
relatively small area.

Several points must be kept in mind when interpreting the images in
Fig.~\ref{fig:maps}. The magnetic field analysis assumes that the spectral
line depth is uniform over the stellar surface and does not take the
presence of dark starspots into account. It is likely that the actual
field in the starspots is much stronger than that reconstructed using ZDI.  
In particular, as the brightness and line depth in the polar region are
suppressed, the field from this region will not contribute to the
circularly polarized spectra. It is possible that there is very strong
radial field in the center of the polar spot and that all we can observe
using the circularly polarized profiles is the brighter penumbral region.
Another point to keep in mind is that the meridional (N-S) component has a
reduced contribution to circularly polarized spectra obtained from high
inclination stars like AB Dor ($i=60^{\circ}$).  This explains why the
field recovered in Fig.~\ref{fig:maps}d is weak compared to the radial and
azimuthal field maps.

\subsection{Coronal extrapolation of the surface field}

\begin{figure}[h]
\includegraphics*[width=17cm]{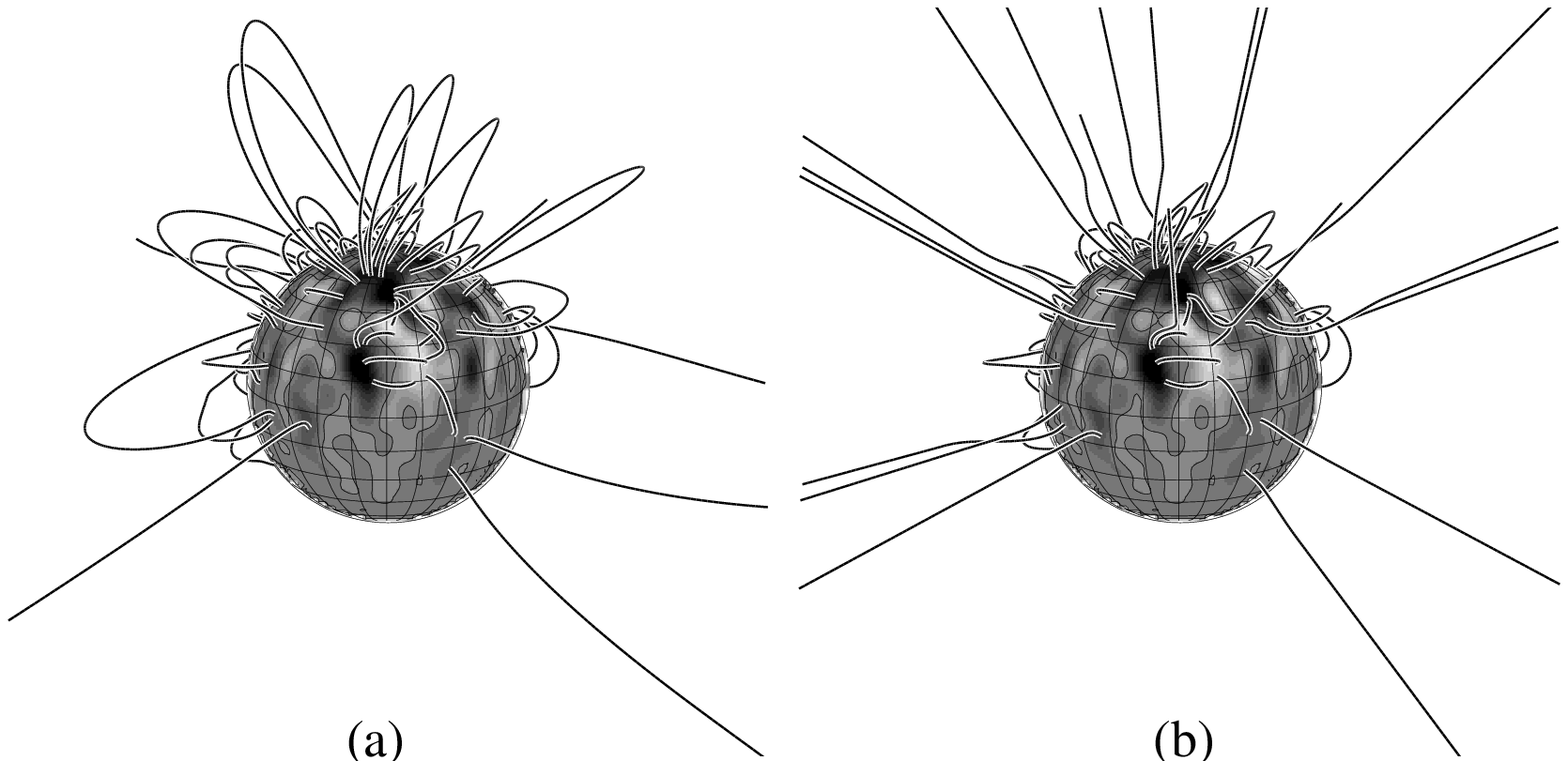}
\caption{Coronal field topology for AB Dor. 
The source surface has been set to a) 5{\,\mbox{$\mbox{R}_*$}};
and b) 1.6{\,\mbox{$\mbox{R}_*$}}.}
\label{fig:corfield}
\end{figure}

The magnetic field, {\bf B}, can be written in terms of spherical harmonic
functions, and the assumption that, $\mbox{\bf j}=-\nabla\Psi$ can be used
to determine the radial dependence of the spherical harmonic coefficients.
The resulting field pattern is shown in Fig.~\ref{fig:corfield}.  
As prominences are observed out to 5{\,\mbox{$\mbox{R}_*$}}\ we initially
computed the field topology by setting the source surface to
5{\,\mbox{$\mbox{R}_*$}} (Fig.~\ref{fig:corfield}a). The source surface is
the point beyond which all the field is assumed to be open. The technique
developed by Jardine et al. (2001) allows us to evaluate locations around
AB Dor that can support slingshot prominences. Applying this technique to
this model shows that the field around AB Dor is complex enough to support
prominences at sufficiently large distances around AB Dor, if the source
surface radius is set to values above 5{\,\mbox{$\mbox{R}_*$}}.
Fig.~\ref{fig:corfield}b shows the field topology for the same model but
with the source surface set to 1.6{\,\mbox{$\mbox{R}_*$}}, which is the
value suggested by our modeling of coronal loops (see below). If this is
the case it is unclear how prominences can be supported.

\section{Heating in stellar corona}

We have used the 3D magnetic field models shown in  
Figs.~\ref{fig:corfield}a~\&~b to study coronal heating in AB Dor. 
In our model we take non-thermal heating to be constant in time and seek
steady state solutions of the energy balance equations.  As the loops are
allowed to be asymmetric, there is in general a steady mass flow along the
loop. The flow velocity is assumed to be smaller than the sound speed, so
the plasma is nearly in hydrostatic equilibrium. We assume that magnetic
flux is constant along the loop. Hence, as the magnetic field strength,
$|B|$, drops off with height, the loop cross-sectional area will expand
(Schrijver et al. 1989).  Energy transport in the lower transition region
($T< 4\times 10^5$~K) is modeled using a parametrization of ambipolar
diffusion. Boundary conditions placed on the footpoints of each loop
include setting the footpoint temperature to $T=2\times 10^4$~K, and
energy fluxes are calculated for each footpoint assuming a simple model of
the chromosphere.

The heating rate is assumed to depend only on the magnetic field strength:
\begin{equation}
E_H(s)=\epsilon_{0} B(s)^{n},
\end{equation}
where $E_H$ is the heating rate per unit volume as function of position, 
$s$, along the loop, and $\epsilon_0$ is a constant.
We find that, if the heating rate is proportional to $B$ ($n=1$), then 
for high-altitude loops $E_H$  
drops off quickly with height and there is insufficient
heating at the apex of the loops for a stable solution. 
Thermal instabilities arise because the heating at the footpoints 
is strong, increasing pressure everywhere in the loop. The heating rate
cannot compensate for local radiative
losses at the top of the loop and this triggers the 
formation of a cool coronal condensation.
Therefore, in the present paper we focus on models with heating independent
of field strength, $n=0$.

In this paper we assume that the EMD peak is caused by loops with a 
maximum temperature $T_{max}\approx 8\times 10^6$~K.
Therefore the constant, $\epsilon_0$, is selected to fit $T_{max}$,
so for short loops $\epsilon_0$ must be very large, and for long loops
$\epsilon_0$ must be very small. The resulting gas pressure is inversely 
proportional to loop length, L. We find that for high-altitude loops
the gas pressure is larger than the magnetic pressure at the apex
of the loop, i.e. the coronal plasma is no longer magnetically contained.
This problem arises for loop heights greater than about 0.6{\,\mbox{$\mbox{R}_*$}},
which suggests that beyond this height all magnetic fields are open.
Therefore, in the following we only present results for the model
with the source surface at 
1.6{\,\mbox{$\mbox{R}_*$}}\
(see Fig.~\ref{fig:corfield}b).

To obtain a clear peak in the EMD it is necessary to assume that the
cross-section varies with position along the loop. By allowing a loop to
expand with height, emission at lower temperatures is suppressed as the
legs of the loops at these temperatures have a smaller cross-section. The
hotter gas at the loop apex fills a larger volume and produces the type of
emission peak that is observed
(Schrijver et al. 1989, Ciaravella et al. 1996). 
Expansion factors ($\Gamma=A_{max}/A_{foot}$) 
of between 5-7 are sufficient to explain
the enhanced peak observed in Fig.~\ref{fig:emdobs}.

\begin{figure}[h]
\includegraphics*[width=14cm]{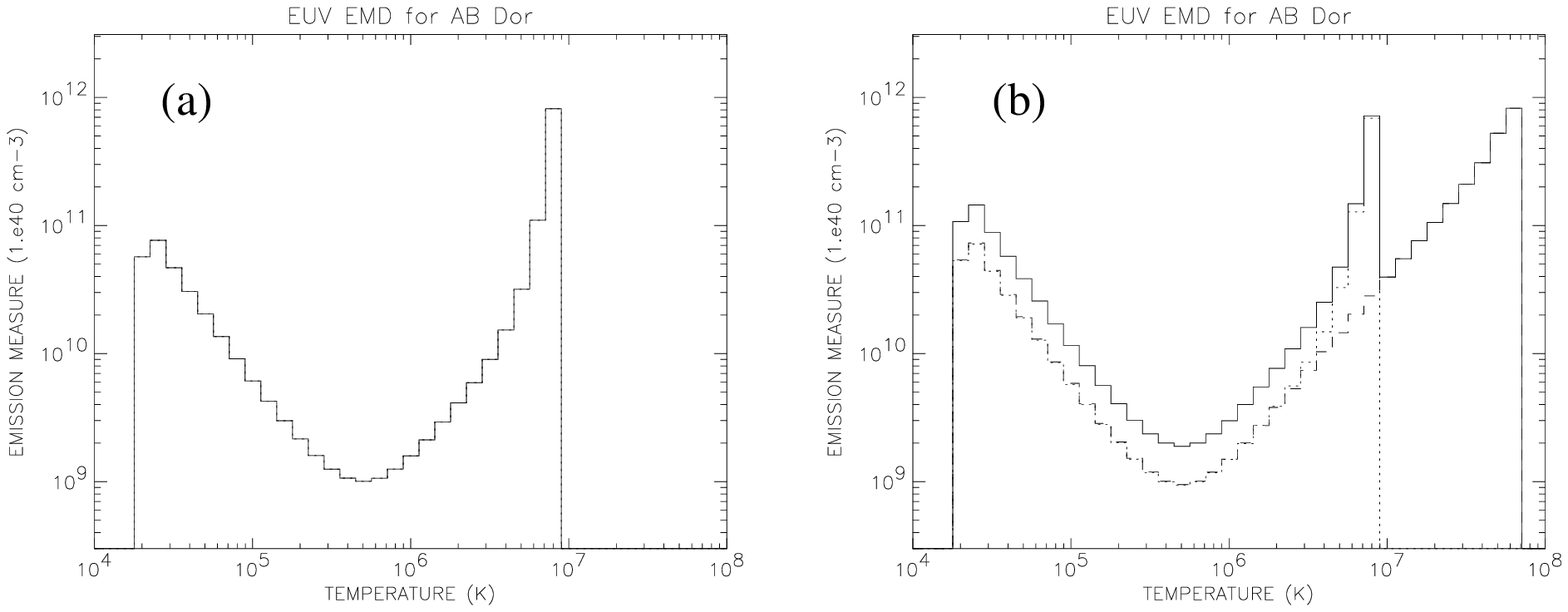}
\caption{EMD for two different models. Each plot shows
how the observed EMD shown at the beginning of this paper can be fit
using loops with 
a) very compact expanding loops ($L\approx 0.0004${\,\mbox{$\mbox{R}_*$}});
and b) a combination of expanding loops
($L\approx0.75${\,\mbox{$\mbox{R}_*$}}; dotted line), 
and more compact dense loops ($L\approx 0.04${\,\mbox{$\mbox{R}_*$}};
dashed line). }
\label{fig:emdfits}
\end{figure}

First, we investigate the type of compact loops that are thought
to be the source of high density measurements at EM peak temperatures 
(Sanz-Forcada et al. 2001, Brickhouse \& Dupree 1998).
We find that, in order to obtain stable loops with coronal density of
about $10^{13}$cm$^{-3}$ as observed by EUVE,  we have to increase the
heating rate such that 
$\epsilon_0\approx 4 \times 10^3$~erg~cm$^{-3}$~s$^{-1}$ 
and we have to reduce the loop length to $L\approx 280$~km 
(we assume $\Gamma=6$).
The gas pressure in such loops would be about $10^4$~dyne/cm$^{2}$ and the 
field strength at the feet of such loops must be at least 3000~G, 
which is larger than the fields observed with ZDI. Assuming there are
about $10^4$ such loops at any one time, we can fit both the observed EMD
(see Fig.~\ref{fig:emdfits}a) and the densities  at the EM peak.
However, their lengths are unrealistic as they are {\em comparable with the
height of the photosphere}.
Furthermore it is unclear why such short loops would have expanding
cross-sections, $ \Gamma > 5$.

Another more realistic model can reproduce the observed EMD but is
inadequate when explaining the high densities in the EM peak.
As mentioned earlier, expansion factors of $5<\Gamma <7$ are needed to 
reproduce the EM peak in Fig.~\ref{fig:emdobs}.
These factors can be obtained using the model shown in  
Fig.~\ref{fig:corfield}b and assuming loop heights in the range
0.1-0.5{\,\mbox{$\mbox{R}_*$}}.
The average length of such a loop is about 0.75{\,\mbox{$\mbox{R}_*$}}\
and if $\epsilon_0 \approx 10^{-3}$~erg~cm$^{-3}$~s$^{-1}$, it will 
have a pressure of about $5$~dyne/cm$^2$, consistent with the TR
observations.
However, the loops have coronal densities of $\approx 3 \times 10^9$~cm$^{-3}$
at $T\approx 8 \times 10^6$~K, much less than implied by the EUVE
observations.
In order to produce higher densities, it is possible to invoke
much smaller loops ($L\approx 0.043${\,\mbox{$\mbox{R}_*$}})
with high heating rates and small expansion factors to fit the high
temperature tail. The EM resulting from a combination of these two types
of loops is plotted in Fig.~\ref{fig:emdfits}b. These hot loops have
densities of $n_e\approx 10^{13}$~cm$^{-3}$ at the EM peak temperature,
but they contribute only 3\% of the emission at this temperature.
Therefore, it is unlikely that this model can reproduce the observed
density sensitive lines.

Both models presented here fit the observed EMD well. The model that can
reproduce the observed densities as well as fit the EM requires loops that
are too short (on the same scale as the height of the photosphere).
The second model would appear more reasonable but it cannot reproduce the
observed density sensitive lines. Therefore, neither model is fully
satisfactory.

\section{Discussion}

Magnetic field maps of AB Dor from 1996 December show very strong
azimuthal field. The non-potential part of the field distribution is
concentrated around the boundary of the dark stable polar spot that
extends down to about 70$^{\circ}$ latitude. This azimuthal field has no
solar analogy but as it is likely that the polar cap is censoring the
field at the pole of the star these maps may not present the full picture.
If strong radial field is concentrated in the most spotted parts of the
star, we may only  be detecting the penumbral regions of the spots located
at the pole.  Schrijver (this volume) describes a model for a spun-up
solar-type star.  According to this model there are three bands of
alternating polarity encircling the pole. If the star is rotating
rapidly, the field lines connecting between the different bands are likely
to be sheared in one direction, thus producing a strong unidirectional
azimuthal field near the pole. However, we find no evidence for bands of
alternating polarity near the pole in our ZDI maps. If they exist,
the polarization signature from the alternating  radial field
bands is buried in the dark polar spot and hence cannot be detected
using our method of magnetic field mapping.

By extrapolating our surface maps, we can model the coronal topology of
the star. These coronal fields can then be used to model heating as well
as to evaluate sites of stable mechanical equilibria that can support
slingshot-type prominences. The fields are sufficiently complex
to support prominences out to the distances at which they are observed
(2-5{\,\mbox{$\mbox{R}_*$}}) assuming the source surface is extending out
this far. It may be necessary to compensate for the missing flux at the
pole and model for the flux we cannot observe in the unseen hemisphere of
AB Dor (stellar inclination=60$^{\circ}$). By adding this missing flux it
is likely that the global pattern will be modified from the type presented
here and that it will become dominated by a dipole field 
(Jardine et al. 2001). 

Using heating rates that can reproduce
the observed transition region densities and pressures, we can fit the
general shape of the observed emission measure distribution. 
The preliminary fits have involved simulating loops with the same general
characteristics as those recovered in our magnetic field model of AB Dor
and computing the filling factors and heating rates required in order to
fit the observed EMD. Jardine et al. (this volume) show how extending the
modeling to the entire corona can be used to evaluate the amount of X-ray
modulation expected. 

In conclusion, we present a loop model that accounts for realistic
asymmetric loop geometries by driving steady mass flows. 
The observed TR densities of $n_e\approx 2-3 \times 10^{12}$~cm$^{-3}$ 
at $T\approx 3 \times 10^4$~K can be reproduced with loops that extend 
to heights of 0.1-0.5{\,\mbox{$\mbox{R}_*$}},
have typical loop lengths of 0.75{\,\mbox{$\mbox{R}_*$}},
and peak temperatures of about $8 \times 10^6$~K.
The higher densities required to fit the EUVE observations remain
difficult to explain satisfactorily. 
If there are loops with a range of heating rates and
densities, as is likely, it may be possible to explain moderately high
densities in the emission peak
in terms of loops with lengths of about 0.75{\,\mbox{$\mbox{R}_*$}}\
($n_e\ge 10^9$~cm$^{-3}$ at $T\approx 8 \times 10^6$~K  ) co-existing with 
much more compact, dense loops ($L\approx 0.04${\,\mbox{$\mbox{R}_*$}},
$n_e\approx 10^{13}$~cm$^{-3}$ at $T\approx  10^7$~K).
However, even on combining these two types of loops, the density 
at the emission peak temperature is still likely to be 
lower than the observed value. 
Once observations of AB Dor taken at different epochs from instruments with
different temperature sensitivities become available, 
we stand to gain more insight into the structures causing these puzzling
density measurements.

The model used to compute the EMD  cannot fully
account for the types of loops that can support  
slingshot prominences out to 4-5{\,\mbox{$\mbox{R}_*$}}. 
It is possible that the corona is largely open beyond about 
1.6{\,\mbox{$\mbox{R}_*$}}\ but that closed field can form at
heights above this where open field lines reverse polarity.
As indicated earlier, coronal condensations tend to form at loop
heights greater than 1.6{\,\mbox{$\mbox{R}_*$}. Could these 
condensations explain the  presence of prominences
at heights up to 5{\,\mbox{$\mbox{R}_*$}} ?

\index{*HD 36705|see {AB Dor}}

\end{document}